\documentclass[prd,eqsecnum,twocolumn,amsfonts,amssymb]{revtex4}

\usepackage{graphicx}

\usepackage{bm}

\newcommand{\beq}{\begin{equation}}
\newcommand{\eeq}{\end{equation}}
\newcommand{\beqs}{\begin{eqnarray}}
\newcommand{\eeqs}{\end{eqnarray}}

\begin{document}

\title{Effect of Scheme Transformations on a Beta Function with Vanishing One-Loop Term}

\author{Thomas A. Ryttov$^a$ and Robert Shrock$^b$}

\affiliation{(a) \ CP$^3$-Origins \\
Southern Denmark University, Campusvej 55, Odense, Denmark}

\affiliation{(b) \ C. N. Yang Institute for Theoretical Physics and
Department of Physics and Astronomy, \\
Stony Brook University, Stony Brook, New York 11794, USA }

\begin{abstract}

  It is commonly stated that because terms in the beta function of a
  theory at the level of $\ell \ge 3$ loops and higher are
  scheme-dependent, it is possible to define scheme transformations
  that can be used to remove these terms, at least in the vicinity of
  zero coupling. We prove that this is not, in general, possible in
  the situation where a beta function is not identically zero but has
  a vanishing one-loop term.

\end{abstract}

\maketitle


\section{Introduction}
\label{intro_section}

Let us consider a quantum field theory in $d$ spacetime dimensions,
with some set of fields and a dimensionless interaction coupling $g$.
Two cases of particular interest are a non-Abelian gauge theory in
$d=4$ dimensions, where $g$ is the gauge coupling, and a scalar theory
with a cubic interaction in $d=6$ dimensions, where $g$ is the
coefficient of the cubic interaction.  Quantum corrections render $g$
dependent upon the Euclidean energy/momentum scale $\mu$ where it is
measured: $g=g(\mu)$.  The dependence of $g$ on $\mu$ is determined by
the renormalization-group (RG) beta function of the theory, $\beta_g =
dg/dt$, where $dt=d\ln\mu$ \cite{rg}. This function has the series
expansion
\beq
\beta_g = g\sum_{\ell=1}^\infty b_\ell a^\ell \ ,
\label{betagseries}
\eeq
where $a = c_d g^2$ \cite{lamform} and $c_d$ is a factor arising from momentum
integrals in loop diagrams: $c_d = S_d/(2\pi)^d$ with $S_d =
2\pi^{d/2}/\Gamma(d/2)$, so, e.g., $c_4=1/(16\pi^2)$, etc. An
equivalent beta function is
\beq
\beta_a = \frac{da}{dt} = 2a \sum_{\ell=1}^\infty b_\ell a^\ell \ .
\label{betaseries}
\eeq
The one-loop coefficient, $b_1$, is independent of the scheme used for
regularization, and (for mass-independent schemes) this is also true
for the two-loop coefficient, $b_2$ \cite{bgz73,gross75}. In contrast,
the higher-loop coefficients $b_\ell$ with $\ell \ge 3$ are
scheme-dependent \cite{msbar}.  It was thus expected that, at least for
sufficiently small coupling, as in the deep ultraviolat (UV) limit of
quantum chromodynamics (QCD), it would be possible to carry out a
scheme transformation that eliminates these terms and yields a beta
function with only one- and two-loop terms \cite{thooft77}.  Such a
scheme has been called the 't Hooft scheme. Refs.
\cite{scc}-\cite{sch3} calculated explicit formulas for the effect of
scheme transformations on the coefficients of a beta function and used
these to construct scheme transformations that remove terms at loop
order $\ell \ge 3$ from the beta function. A set of necessary
conditions for a scheme transformation to be physically admissible was
given in \cite{scc,sch}, and it was shown that although these
conditions can easily be satisfied if one applies a scheme
transformation in the vicinity of the origin in coupling constant
space, as in applications to optimized schemes in perturbative quantum
chromodynamics \cite{brodsky}, they are not automatically satified,
and are a significant constraint on the application of scheme
transformations, when investigating a zero of the beta function away
from the origin.  In an asymptotically free theory, such a zero would
be an infrared fixed point (IRFP) of the renormalization group, while
in an infrared-free theory, such a zero would be an ultraviolet fixed
point (UVFP) of the RG.  An example of an IRFP occurs in an
asymptotically free non-Abelian gauge theory with sufficiently many
massless fermions \cite{bz}, while an (exactly solved) example of a
UVFP occurs in the O($N$) nonlinear  $\sigma$ model in
$d=2+\epsilon$ dimensions in the large-$N$ limit \cite{nlsm}.

In this work we consider the situation in which a theory has a beta
function (which is not identically zero) with vanishing one-loop
coefficient, $b_1=0$.  We prove that in this case it is not, in
general, possible to construct and apply a scheme transformation, even
in the vicinity of the origin, $a=0$, that removes the
scheme-dependent terms $b_\ell$ with $\ell \ge 3$ in the beta
function.  We discuss implications for the study of zeros of the beta
function in this type of theory. 


%
\section{Scheme Transformations}
\label{st_section}

In this section we give a brief review of relevant methodology on
scheme transformations that will be needed here.  One may define a
scheme transformation as a mapping relating $a$ and $a'$ given by
\beq
a = a' f(a') \ ,
\label{aap}
\eeq
where $f(a')$ is the scheme transformation function.  If the theory is free,
then its properties must remain the same under a scheme transformation, so
$f(0)=1$. The function $f(a')$ is taken to have the power series expansion
\beq
f(a') = 1 + \sum_{s=1}^{s_{max}} k_s (a')^s \ ,
\label{fapseries}
\eeq
where $s_{max}$ may be finite or infinite. The corresponding Jacobian
$J=da/da'$ has the series expansion 
\beq
J = 1 + \sum_{s=1}^{s_{max}} (s+1)k_s (a')^s \ .
\label{j}
\eeq

The beta function in the transformed scheme is
\beq
\beta_{a'} \equiv \frac{da'}{dt} = \frac{da'}{da} \, \frac{da}{dt} =
J^{-1} \, \beta_a \ .
\label{betaap}
\eeq
This beta function has the series expansion
\beq
\beta_{a'} = 2a' \sum_{\ell=1}^\infty b_\ell' (a')^\ell
\label{betaprime}
\eeq
with a new set of coefficients $b_\ell'$.  As noted above, the
one-loop and two-loop coefficients are left invariant by this scheme
transformation, i.e., $b_1'=b_1$ and $b_2'=b_2$
\ \cite{bgz73,gross75}.  Ref. \cite{scc} presented explicit
expressions for the $b_\ell'$ in terms of the $b_\ell$ and $k_s$ for
loop order $3 \le \ell \le 5$ and Ref. \cite{sch} extended these up to
$\ell=8$ inclusive. For $b_3'$ and $b_4'$ these expressions are
\cite{scc}
\beq
b_3' = b_3 + k_1b_2+(k_1^2-k_2)b_1
\label{b3prime}
\eeq
and 
\beq
b_4' = b_4 + 2k_1b_3+k_1^2b_2+(-2k_1^3+4k_1k_2-2k_3)b_1 \ .
\label{b4prime}
\eeq
For the reader's convenience, we list some of these
expressions for $b_\ell'$ with higher $\ell$ in Appendix
\ref{bellprime_general}, from Refs. \cite{scc,sch}.

As noted above, one important application of the study of schemes
and scheme transformations is to calculations of higher-order
corrections in perturbative QCD scattering processes at
high energies.  In this application, one is interested in choosing a
scheme such that higher-order terms are small, so that one
can achieve as accurate as possible a description of experimental
data at a given order in perturbation theory. A different type
of application is to the investigation of a possible zero of the
beta function away from the origin in coupling constant space.

As specified in \cite{scc,sch}, in order for a scheme transformation
to be physically acceptable, it must satisfy the following necessary
conditions: 
\begin{itemize}

\item

C$_1$: the scheme transformation must map a real positive $a$ to a real
positive $a'$, since a map taking $a > 0$ to $a'=0$ would be
singular, and a map taking $a > 0$ to a negative or complex $a'$
would violate the unitarity of the theory.

\item

C$_2$: the scheme transformation should not map a
value of $a$ for which perturbation theory may be reliable, to a
value of $a'$ that is so large that perturbation theory is
unreliable.

\item

C$_3$: \ $J$ should not vanish in the region of $a$ and
$a'$ of interest, or else there would be a pole in Eq. (\ref{betaap}).

\item

  C$_4$: \quad The existence of an IR or UV zero of $\beta$ has physical
  significance and must therefore be scheme-independent. 
  Hence, a scheme transformation must satisfy the condition that
  $\beta_a$ has a zero away from the origin if and only if
$\beta_{a'}$ has a corresponding zero away from the origin.

\end{itemize}
Clearly, these conditions apply both for a given scheme transformation and
its inverse. 

Although $b_1$ is nonzero in QCD, there are theories in which
$b_1$ may be zero (without the beta function being identically zero).
One example of a theory in which $b_1$ can vanish is a vectorial
non-Abelian gauge theory with gauge group $G$ and $N_f$ Dirac fermions
transforming according to a representation $R$ of $G$. The one-loop
coefficient of the beta function is \cite{b1}
\beq
b_1 = -\frac{1}{3}( 11 C_A - 4N_f T_f) \ , 
\label{b1}
\eeq
and the two-loop coefficient is \cite{b2}
\beq
b_2 = -\frac{1}{3}\Big [ 34C_A^2 - 4(5C_A+3C_f)N_f T_f \Big ] \ , 
\label{b2}
\eeq
where $C_A$ and $C_f=C_2(R)$ are the quadratic Casimir invariants of
the adjoint representation and the fermion representation $R$,
respectively, and $T_f=T(R)$ is the trace invariant of $R$
\cite{group_invariants}.  The coefficient $b_1$ vanishes if
$N_f=N_{f,b1z}$, where
\beq
N_{f,b1z} = \frac{11C_A}{4T_f} \ .
\label{Nfb1z}
\eeq
If $N_f=N_{f,b1z}$, then $b_2 = C_A(7C_A+11C_f)$, which is positive,
so the theory with $N_f=N_{f,b1z}$ is IR-free.  As an explicit
example, one could take $G={\rm SU}(2)$ and $R$ equal to the
fundamental representation, so that $N_{f,b1z}=11$. That is, an SU(2)
gauge theory with these 11 Dirac fermions has $b_1=0$. Examples can also
be given of chiral gauge theories in which, for a special choice of
gauge group and fermion content, $b_1=0$.  In all of these cases,
the choice of parameters that renders $b_1=0$ leaves a nonzero $b_2$.

Moreover, the vanishing of $b_1$ can occur in scalar field theories; a
recent example is a scalar theory with a cubic self-interaction in
$d=6$ dimensions in which the scalar transforms as a bi-adjoint
representation of a global ${\rm SU}(N) \otimes {\rm SU}(N)$ symmetry
group \cite{gracey2020}.  In this theory, the first nonzero term
in the beta function is $b_2$, which is negative. A study of a
possible IRFP in this theory was carried out in \cite{phi36}.

A different type of situation occurs in an ${\cal N}=2$ supersymmetric
gauge theory with gauge group SU($N_c$) and $N_f$ matter
hypermultiplets. A closed form expression for the beta function was
calculated in \cite{nsvz} (see also \cite{ss}). This beta
function has the property that $b_n = 0$ if $n \ge 2$.  By choosing
$N_f$ appropriately, one can make $b_1=0$, so that the beta function
vanishes identically.  In contrast, here we discuss theories in which
$b_1=0$ (either because of a special choice of parameters, as in
Eq. (\ref{Nfb1z}), or identically, as in \cite{gracey2020,phi36}), but
the beta function is not identically zero.


\section{Schemes to Remove Terms in Beta Function of Order
  Three Loops and Higher}
\label{removal_section}

An important application of scheme transformations is to the
analysis of possible zero(s) of the beta function away from the
origin.  The beta function of an asymptotically free non-Abelian gauge
theory has an ultraviolet zero at $a=\alpha/(4\pi)=0$, which is a
UVFP.  If the theory contains sufficiently many massless fermions, the
(perturbatively calculated) beta function may also have an infrared
zero at a nonzero value of the gauge coupling.  The theory is
weakly coupled at this IRFP if the number of fermions is close to the
upper limit allowed by asymptotic freedom, namely $N_{f,b1z}$ in a
theory with fermions in a single representation $R$, and hence is
amenable to a perturbative treatment using series expansions in the
variable $\Delta_f = N_u-N_f$ \cite{bz}.

Since the terms of loop order $\ell \ge 3$ in the beta function are
scheme-dependent, so is the value of the IR zero when calculated to
three-loop or higher-loop order.  In order to understand the physical
implications of this IR zero, it is necessary to assess the effect of
scheme dependence on its value.  A study of this dependence was
carried out in \cite{sch,sch2} using several scheme transformations. 
Related studies were performed in \cite{trs}-\cite{gracey_simms}.

One type of procedure that would be natural for a quantitative study
of scheme-dependence of a zero of the beta function would be to
construct and apply a scheme transformation that would remove
successively higher-loop terms in the beta function and, at each
stage, determine how this removal shifted the position of the IR zero.
Extending the results of \cite{sch}, Ref. \cite{sch2} constructed a
set of scheme transformations $S_{R,m}$ with $m \ge 2$ with $k_1=0$ in
Eq. (\ref{fapseries}) that remove the terms in the beta function at
loop order $\ell=3$ to $\ell=m+1$, inclusive and determined the range
of $\alpha$ over which $S_{R,2}$ and $S_{R,3}$ could be applied to
study the IR zero of the beta function of an asymptotically free gauge
theory while satisfying the criteria to avoid introducing unphysical
pathologies.  Ref. \cite{sch3} presented a generalized one-parameter
class of scheme transformation, denoted $S_{R,m,k_1}$ with $m \ge 2$,
depending on $k_1$, with the property that an $S_{R,m,k_1}$ scheme
transformation eliminates the $\ell$-loop terms in the beta function
of a quantum field theory from loop order $\ell=3$ to order
$\ell=m+1$, inclusive.  A transformation in this class reduces to
$S_{R,m}$ if $k_1=0$. These types of scheme transformations have also
been used in the analysis of a possible ultraviolet zero in the beta
functions of a U(1) gauge theory \cite{lnf} and an O($N$) $\lambda
|\vec \phi |^4$ theory \cite{lam}.

Although our focus here is on scheme transformations,
we note that one can also analyze properties of (physical, gauge-invariant)
operators at the IRFP in a non-Abelian gauge theory with sufficiently many
fermions via series expansions in powers of $\Delta_f$.  These have the
advantage that they are manifestly scheme-independent.  This program has
been carried out up to the $O((\Delta_f)^3)$ level in \cite{gtr,dex} and
up to the $O((\Delta_f)^4)$ level in \cite{gsi,dexl}, the latter using the
five-loop beta function \cite{b5su3,b5}. 

To set the stage for our new results, we briefly recall the procedure for
the construction of the $S_{R,m,k_1}$ scheme transformation in \cite{sch3}. 
The first step is to use Eq. (\ref{b3prime}) and solve the equation
$b'_3=0$ for $k_2$. This yields the result
\beq
k_2 = \frac{b_3}{b_1} + \frac{b_2}{b_1} \, k_1 + k_1^2 \quad {\rm for} \
S_{R,m,k_1} \ {\rm with} \ m \ge 2 \ .
\label{k2solk1}
\eeq
This suffices for $S_{R,2,k_1}$. The reason that we solve for $k_2$ instead
of $k_1$ is that this involves the solution of a linear equation for $k_1$,
whereas the equation $b_3'=0$ is a quadratic equation in $k_1$, so one would
have to choose which of the two solutions of this quadratic would be used.

To obtain $S_{R,m,k_1}$ with $m \ge 3$, removing the $\ell=3, \ 4$
terms in $\beta_{a'}$, one substitutes the solution for $k_2$
from Eq. (\ref{k2solk1}) into Eq. (\ref{b4prime}) and solves the
equation $b'_4=0$ for $k_3$.  Again, this is a linear equation, with a
unique solution, which is
\beq
k_3 = \frac{b_4}{2b_1} + \frac{3b_3}{b_1}\, k_1 + \frac{5b_2}{2b_1}\, k_1^2
+ k_1^3 \quad {\rm for} \ S_{R,m,k_1} \ {\rm with} \ m \ge 3 \ .
\label{k3solk1}
\eeq
One continues in this manner to determine the $k_s$ with $s \ge 4$ such
as to remove the terms in the beta function up to successively higher
loop orders.

This procedure requires $b_1$ to be nonzero, since
otherwise various multiplicative factors involving $1/b_1$ in the
$k_s$ with $s \ge 2$ and higher powers of $1/b_1$ in the $k_s$ with $s
\ge 4$ would be singular.  Here we investigate the situation where
$b_1=0$. 


\section{Impossibility of Removing All Higher-Loop Terms with $\ell \ge 3$
if $b_1=0$: Case where $b_2 \ne 0$}
\label{stb1z_section}

In this and the next section we show that if the one-loop term in the
beta function is zero, i.e., if $b_1=0$, then it is, in general, not
possible to apply Eqs. (\ref{aap}) and (\ref{fapseries}) to transform
to a scheme in which all of the scheme-dependent $\ell$-loop
coefficients with $\ell \ge 3$ are zero.  We begin in this section
with the case where $b_1=0$ and $b_2 \ne 0$.  If $b_2 > 0$, then this
theory is IR-free, while if $b_2 < 0$, the theory is UV-free (i.e.,
asymptotically free).

We proceed to analyze scheme transformations intended to try to
set higher-loop coefficients equal to zero.  From Eq. (\ref{b3prime}),
it follows that in order to have $b_3'=0$, the unique solution for
$k_1$ in the scheme transformation (\ref{aap})-(\ref{fapseries}) is
\beq
k_1 = -\frac{b_3}{b_2} \ .
\label{k1solb1z}
\eeq
Substituting this in Eq. (\ref{b4prime}) for $b_4'$, we obtain
\beq
b_4' = b_4 - \frac{b_3^2}{b_2} \ . 
\label{b4primeb1z}
\eeq
In general, this is nonzero. This proves that if $b_1=0$, then there
is, in general, no scheme transformation of the form (\ref{aap}) with
(\ref{fapseries}) that can be used to render $b_4'$ zero, and hence,
{\it a fortiori}, it is not possible to remove all of the
scheme-dependent terms in the beta function.

For completeness, we comment on the extent to which one can remove
higher-loop terms with $\ell \ge 5$ in this case. Substituting
the value of $k_1$ from (\ref{k1solb1z}) in Eq. (\ref{b5prime}) for
$b_5'$, we obtain 
\beq
b_5' = b_5 -\frac{3b_3b_4}{b_2} + \frac{3b_3^3}{b_2^2} -2k_2b_3-k_3b_2 \ .
\label{b5primeb1z}
\eeq
It is always possible to render this $b_5'=0$ by setting
\beq
k_3 = \frac{b_5}{b_2} -\frac{3b_3b_4}{b_2^2} + \frac{3b_3^3}{b_2^3} -\frac{2k_2b_3}{b_2} \ .
\label{k3solb1z}
\eeq
As is evident from Eq. (\ref{k3solb1z}), there is thus an infinite set of pairs
$(k_2,k_3)$ that render $b_5'=0$. Note that in the special case where $b_3=0$,
$k_3$ takes on the unique value $k_3 = b_5/b_2$ while $k_2$ is undetermined.
Examining Eq. (\ref{b6prime}), we see that with $k_3$ given by
Eq. (\ref{k3solb1z}), one can solve the equation $b_6'=0$ (as a linear
equation in $k_4$) for a value of $k_4$.  Similarly, with these $k_s$
values chosen, one can solve $b_7'=0$ (as a linear equation in $k_5$)
for $k_5$, and so forth for higher $b_\ell'$.

Thus, in the transformed scheme,
\beqs
\beta_{a'} &=& 2(a')^3[b_2 + b'_4 (a')^2] \cr\cr
&=& 2(a')^3\Big [ b_2 + \Big ( b_4 - \frac{b_3^2}{b_2} \Big )(a')^2 \Big ] .  
\eeqs
The function $\beta_{a'}$ has a formal zero away from the origin at
$a=a_z$, where 
\beq
a'_{z} = \bigg [ -\frac{b_2}{b_4 - \frac{b_3^2}{b_2} } \bigg ]^{1/2} \ .
\label{asolb1z}
\eeq
This is physical if the expression in the square root is positive. There are two cases
with nonzero $b_2$ to consider, namely $b_2 < 0$ and $b_2 > 0$.  Let us first consider
the case $b_2 < 0$, where the theory is UV-free.  Then the condition that $a'_z$ is
physical is that that $b_4 - (b_3^2/b_2) > 0$, i.e.,
\beq
b_4 > -\frac{b_3^2}{|b_2|}  \quad {\rm for \ IRFP \ if} \ b_1=0 \ {\rm and} \ b_2 < 0 \ . 
\label{asolb1z_irfp_b4condition}
\eeq
This is a necessary condition for the theory to have a physical IRFP at $a'_z$,
but is not sufficient; the scheme transformation to the primed scheme must also
satisfy the conditions C$_1$ - C$_4$ from \cite{scc,sch} listed above,

Next, we consider the case $b_2 > 0$, where the theory is IR-free.
Here the condition that $a'_z > 0$ is that $b_4-(b_3^2/b_2) < 0$,
i.e.,
\beq
b_4 < \frac{b_3^2}{b_2} \quad {\rm for \ UVFP \ if} \ b_1=0 \ {\rm and} \ b_2 > 0 \ . 
\label{asolb1z_uvfp_b4condition}
\eeq
Again, this is a necessary but not sufficient condition for the theory to a physical
UVFP at $a'_z$; the scheme transformation must also satisfy conditions C$_1$ - C$_4$.


\section{Impossibility of Removing All Higher-Loop Terms with $\ell \ge 3$
if $b_1=0$: Case Where $b_2=0$}
\label{stb1b2z_section}

Next, we consider the case in which both $b_1=0$ and $b_2=0$, i.e., the maximal
scheme-independent part of the beta function is zero. Here, Eq. (\ref{b3prime})
reduces to
\beq
b_3'=b_3 \ .
\label{b3prime_b1b2z}
\eeq
In the generic situation in which $b_3 \ne 0$, this immediately proves that
if $b_1=b_2=0$, then there is, in general, no scheme transformation that one can
use to remove all higher-loop terms with $\ell \ge 3$ in the beta function.
Here we assume that $b_3 \ne 0$ and comment on the special case $b_3=0$ below. 
We remark on specific results for other higher-loop coefficients.
The condition that $b_4'=0$ can be satisfied by choosing
\beq
k_1 = -\frac{b_4}{2b_3} \ .
\label{k1sol_b1b2z}
\eeq
Substituting this into Eq. (\ref{b5prime}), we see that it is possible to render
$b_5'=0$ with the choice
\beq
k_2 = -\frac{b_5}{b_3} + \frac{b_4^2}{b_3^2} \ .
\label{k2sol_b1b2z}
\eeq
Substituting these values of $k_1$ and $k_2$ into the expression (\ref{b6prime}), we
obtain
\beq
b_6' = b_6 -\frac{2b_4b_5}{b_3} + \frac{b_4^3}{b_3^2} \ .
\label{b6prime_b1b2z}
\eeq
This does not contain dependence on any other $k_s$ that can be chosen
to make it zero, and, in general, it is nonzero.  With $k_1$ and $k_2$
set equal to their values in Eqs. (\ref{k1sol_b1b2z}) and
(\ref{k2sol_b1b2z}), there is an infinite number of values of the pair
$(k_3,k_4)$ that can be used to render $b_7'=0$.  Having done this,
one can choose $k_5$ to make $b_8'=0$.  Higher-loop coefficients can
be analyzed in a similar manner.

We next consider the hypothetical case in which not only $b_1=0$ and
$b_2=0$, but also one starts in a scheme in which some finite number of
higher-loop coefficients $b_\ell$ with $3 \le \ell \le p$ are zero. As before,
we find that it is not, in general, possible to construct a scheme
transformation that renders all of the $b_\ell$ with $\ell \ge p+1$
zero. This is simply proved by noting that our general results in
Refs. \cite{scc,sch} have the form
\beq
b_\ell' = b_\ell + (\ell-2)k_1b_{\ell-1} + ... \ {\rm for} \ \ell \ge 3, 
\label{bellprimeform}
\eeq
where the $...$ in Eq. (\ref{bellprimeform}) denote a sum of
$k_s$-dependent coefficients times the coefficients $b_k$ with $1 \le
k \le \ell-2$.  Hence, if $b_\ell=0$ for $1 \le \ell \le p$, then
$b_{p+1}'=b_{p+1}$.  Since, by assumption, $b_{p+1}$ is nonzero, so is
$b_{p+1}'$, which proves our claim.


%
\section{Conclusions}
\label{conclusion_section}

In conclusion, in this work we have proved that if the beta function of
a theory is not identically zero and if the one-loop term
in this beta function is zero, then, in general, it is not possible to
transform to a scheme where all of the scheme-dependent coefficients
$b'_\ell$ with $\ell \ge 3$ are zero.  We have given explicit results
for several specific cases, including the case in which $b_1=0$ but
$b_2 \ne 0$ and the case where $b_1=b_2=0$.  In the case of greatest
physical interest, namely where $b_1=0$ and $b_2 \ne 0$, we have also
discussed resultant necessary (but not sufficient) conditions for the
existence of a zero in the beta function away from the origin.


\begin{acknowledgments}

  We thank J. Gracey for valuable discussions in connection with
  \cite{gracey2020} and collaboration on \cite{phi36}.  This research
  was supported in part by the U.S. National Science Foundation Grant
  NSF-PHY-1915093 (R.S.).

\end{acknowledgments}


\bigskip
\bigskip

\begin{appendix}

\section{Expressions for the Beta Function Coefficients $b_\ell'$}
\label{bellprime_general}

For reference, we list some of the higher-loop coefficients $b_\ell'$
with $\ell \ge 5$ from Refs. \cite{scc,sch}.
\begin{widetext}
\beq
b_5' = b_5+3k_1b_4+(2k_1^2+k_2)b_3+(-k_1^3+3k_1k_2-k_3)b_2
     + (4k_1^4-11k_1^2k_2+6k_1k_3+4k_2^2-3k_4)b_1
\label{b5prime}
\eeq
\beqs
b_6' & = & b_6 +4k_1b_5+(4k_1^2+2k_2)b_4+4k_1k_2b_3
 + (2k_1^4-6k_1^2k_2+4k_1k_3+3k_2^2-2k_4)b_2 \cr\cr
    & + & (-8k_1^5+28k_1^3k_2-16k_1^2k_3-20k_1k_2^2
      + 8k_1k_4+12k_2k_3 -4k_5)b_1 
\label{b6prime}
\eeqs
and
\beqs
b_7' & = & b_7 +5k_1b_6+(7k_1^2+3k_2)b_5+
     (2k_1^3+7k_1k_2+k_3)b_4 + (k_1^4-2k_1^2k_2+4k_1k_3+3k_2^2-k_4)b_3 \cr\cr
     & + & (-4k_1^5+15k_1^3k_2-9k_1^2k_3-12k_1k_2^2+9k_2k_3
+5k_1k_4-3k_5)b_2 \cr\cr
     & + & (16k_1^6-68k_1^4k_2+40k_1^3k_3-21k_1^2k_4
+73k_1^2k_2^2-58k_1k_2k_3+10k_1k_5+16k_2k_4-12k_2^3+9k_3^2-5k_6)b_1 \ .
\cr\cr
& &
\label{b7prime}
\eeqs
\end{widetext}

\end{appendix} 



\begin{thebibliography}{99}
\bibitem{lamform} For a scalar theory in $d=4$ dimensions with a
  $\lambda \phi^4$ interaction, Eq. (\ref{betagseries}) would be
  $\beta_\lambda = \lambda \sum_{\ell=1}^\infty b_\ell a^\ell$ with
  $a = \lambda/(16\pi^2)$. Our conclusions apply to all of these theories,
  but we use notation specific to the $d=4$ gauge theory and $d=6$
  cubic scalar theory.
\bibitem{group_invariants} We recall the general definitions of these
  group invariants. Denote $T_R$ as a generator of the Lie algebra of
  a group $G$ in the representation $R$. Then the quadratic Casimir
  invariant $C_2(R)$ is defined by $T^a_RT^a_R = C_2(R) I$, where $I$
  is the $d_R \times d_R$ identity matrix, and the trace invariant
  $T(R)$ is defined by ${\rm Tr}_R(T^a_R T^b_R) = T(R)\delta^{ab}$,
  where $1 \le a \le o(G)$, with $o(G)$ the order of the group.  We
  write $C_A=C_2(Adj)$ and $C_f = C_2(R_f)$.  For SU($N_c$),
  $C_2(Adj)=T(Adj)=N_c$ and for the fundamental representation, $C_f =
  (N_c^2-1)/(2N_c)$ and $T_f=1/2$.
\end{thebibliography}
\end{document}